\documentstyle[twoside,fleqn,npb,epsfig]{article}

\newcommand{\AmS}{{\protect\the\textfont2
  A\kern-.1667em\lower.5ex\hbox{M}\kern-.125emS}}

\title{Transverse Double-Spin Asymmetries for Dimuon Production in
$pp$ Collisions}

\author{O. Martin\address{Institut f\"ur Theoretische Physik,
        Universit\"at Regensburg, \\
        D-93040 Regensburg, Germany}%
        \thanks{This work was supported by the BMBF and the ``Deutsche
                Forschungsgemeinschaft''.}}

\begin{document}

\begin{abstract}
We calculate the transverse double-spin asymmetry for the production
of dimuons in $pp$ collisions as function of the dimuon rapidity and
mass to next-to-leading order accuracy in the strong coupling constant.
Predictions for {\sc Bnl-Rhic} and {\sc Hera}-$\vec{N}$ 
are made by assuming a saturation of Soffer's inequality
at a low hadronic input scale. It seems unlikely that transversity can be
measured in dimuon production at {\sc Rhic}.
\end{abstract}

\maketitle

\section{TRANSVERSITY}

\begin{figure}[htb]
\vspace{9pt}
\epsfig{file=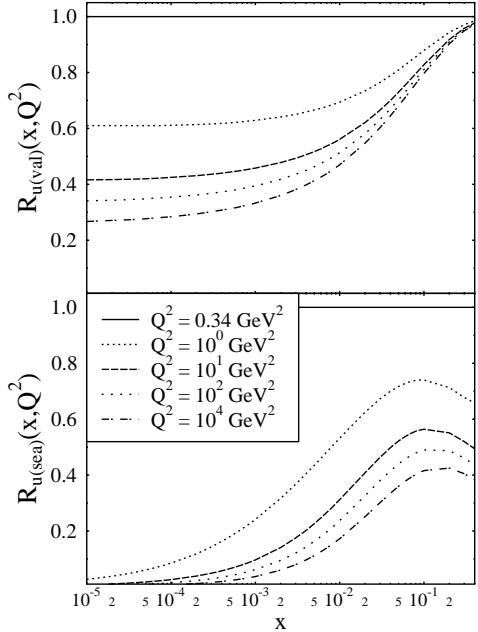,width=6.5cm}
\caption{Preservation of Soffer's inequality by NLO evolution. $R$ is
defined as $R(x,\mu^2)=2 |\delta q(x,\mu^2)|/\left[
q(x,\mu^2)+\Delta q(x,\mu^2)
\right].$}
\label{fig1}
\end{figure}

The~transversity~distribution $\delta q(x,\mu^2)$
counts the number of quarks in a transversely polarized proton weighted
with their transverse polarization \cite{ralston}. 
As a twist-2 distribution
function it is, at least theoretically, as important as the unpolarized and
helicity weighted quark distributions $q(x,\mu^2)$ and $\Delta q(x,\mu^2)$
\cite{jaffeji}.
All three objects are related by Soffer's inequality 
\cite{soffer} which states that for quarks and antiquarks
\begin{equation}
\label{soffersinequality}
|\delta q(x,\mu^2)| \leq \frac{1}{2} \left[q(x,\mu^2)+\Delta q(x,\mu^2)\right]\,.
\end{equation}
This relation is preserved by next-to-leading order (NLO) evolution which means
that it will be valid for all $\mu>\mu_0$ in case it is fulfilled at the scale
$\mu_0$ \cite{preserve}.
 This is demonstrated in Fig.~\ref{fig1} where we assume that Soffer's
inequality is saturated at the hadronic scale of $\mu_0\approx\cal{O}$(0.6 GeV).
Due to lack of data on $\delta q(x,\mu^2)$ all predictions in this work are 
based on this assumption. We use the unpolarized GRV~95~HO 
and the GRSV standard helicity weighted parton distributions.

\section{MUON PAIR PRODUCTION}
As usual, the polarized cross section for the production of 
dimuons is given by a convolution of a partonic
cross section ${\mathrm d}\delta\hat\sigma$ with polarized parton
densities \cite{vowe}:
\begin{eqnarray}
\label{xsec}
&&\frac{{\mathrm d}\delta\sigma}{{\mathrm d}M{\mathrm d}y{\mathrm d}\phi}=
\int_{x_1^0}^1{\mathrm d}x_1\int_{x_2^0}^1{\mathrm d}x_2
\frac{{\mathrm d}\delta \hat\sigma}{{\mathrm d}M{\mathrm d}y{\mathrm d}\phi}\nonumber\\
&&\times\sum_q \bar{e}_q^2\left[\delta q(x_1,\mu_F^2)\delta \bar q(x_2,\mu_F^2) + 
1\leftrightarrow 2\right]\,.
\end{eqnarray}
The dimuon mass $M$ and rapidity $y$ are related to the integration limits
by $x_{1,2}^0=(M/\sqrt{S})\exp{(\pm y)}$, where $S$ is the hadronic center-of-mass 
(cm) energy.
At leading order (LO), $x_{1,2}^0$ are just the momentum fractions of the interacting
(anti-)quarks so that the $y$-dependent cross section is sensitive to the shape
of the transversity densities. We calculated the partonic cross section to NLO 
($\cal{O}$($\alpha_s$)) in the $\overline{\mbox{MS}}$-scheme by transformation of the
results of \cite{vowe} 
which were previously calculated in a massive gluon scheme. The partonic
cross section retains its $\cos{(2\phi)}$-dependence also at NLO with $\phi$ being the
\linebreak
azimuthal angle of one of the outgoing muons. So in order to maximize statistics
we define the transverse double-spin asymmetry as \cite{paper}
\begin{eqnarray}
A_{TT}(M,y) \equiv 
\frac{
\int_{0}^{2\pi}
{\mathrm d}\phi
\,\,{\mathrm d}\delta\sigma(\cos{2\phi}\rightarrow |\cos{2\phi}|)}
{\int_0^{2\pi}{\mathrm d}\phi 
\,\,{\mathrm d}\sigma}\,.\nonumber
\label{att}
\end{eqnarray}

\section{RESULTS}
\begin{figure}[htb]
\vspace{9pt}
\epsfig{file=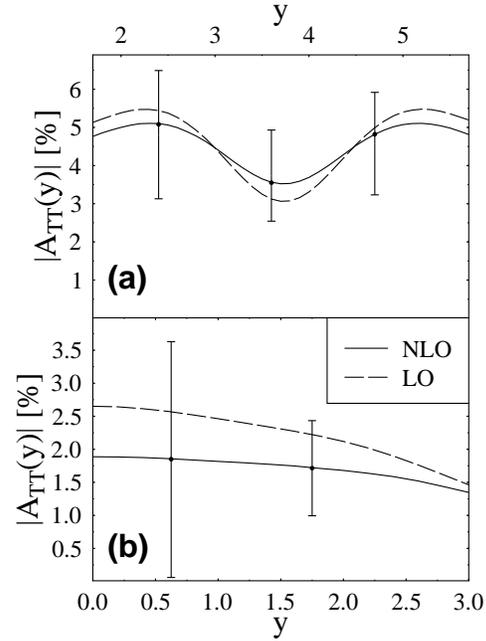,width=6.5cm}
\caption{Rapidity dependence of $A_{TT}$ for (a) {\sc Hera}-$\vec N$ 
($E_{\rm beam}=820$~GeV, $M=4-9$~GeV) \cite{nowak} and (b) {\sc Phenix} at {\sc Rhic} 
($\sqrt{S}=200$~GeV, $M=5-9$~GeV) \cite{phenix}.}
\label{fig2}
\end{figure}

\begin{figure}[htb]
\vspace{9pt}
\epsfig{file=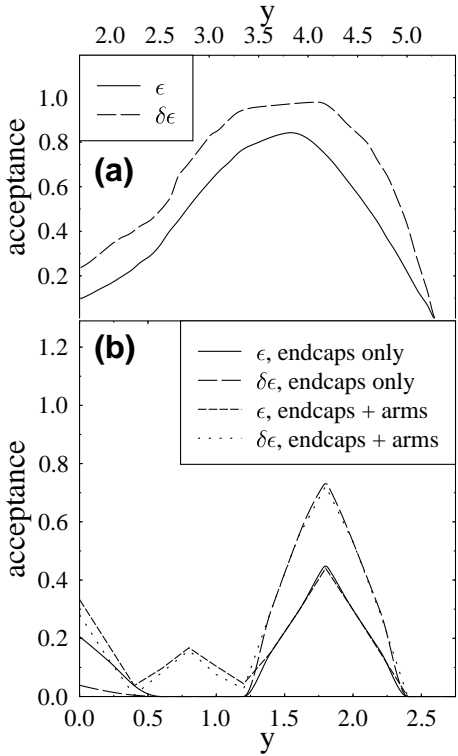,width=6.5cm}
\caption{Unpolarized and polarized acceptance $\epsilon$ and $\delta\epsilon$ 
as function of dimuon rapidity for (a) {\sc Hera}-$\vec N$ and
(b) {\sc Phenix} at {\sc Rhic} (same parameters as in Fig.~2). 
For the geometrical and kinematical cuts see \cite{paper}.}
\label{fig3}
\end{figure}

Fig.~\ref{fig2} shows the $y$-dependence of $A_{TT}$ for {\sc Hera}-$\vec N$ 
at $\sqrt{S}\approx 40$~GeV and 
{\sc Rhic} at $\sqrt{S}=200$~GeV integrated over a suitable dimuon mass range.
QCD corrections to the asymmetry are obviously larger for higher cm-energies while
the opposite is true for the polarized cross sections \cite{paper}. 
What is actually shown is the maximal absolute $A_{TT}$ which is consistent with
the validity of Eq.~(\ref{fig1}) at $\mu\approx 0.6$~GeV. If the inequality 
is not saturated at this scale, then the resulting asymmetry will be smaller.
The statistical errors shown in Fig.~\ref{fig2} are based on an integrated
luminosity of ${\cal L}=240\,{\rm pb}^{-1}$
for {\sc Hera}-$\vec N$, 
${\cal L}=320\,{\rm pb}^{-1}$ for {\sc Rhic} and a beam and target polarization of 
${\cal P}=70$\%.
Furthermore we also took the limited geometrical acceptance
$\epsilon$ and $\delta\epsilon$ of the detectors into
account (cf. Fig.~\ref{fig3}).
For their calculation the momenta of the outgoing
muons must be known which can not be inferred from the variables
$M$, $y$ and $\phi$ of Eq.~(\ref{xsec}). Actually, 
only the square root of the acceptance
enters the expression for the statistical error
\begin{equation}
\label{staterr}
\mbox{stat. error}=\frac{1}{{\cal P}^2\sqrt{{\cal L}\int\epsilon{\mathrm d}\sigma}}\quad,
\end{equation}
and so we just restricted ourselves to a LO calculation since here the transverse
momenta $k_T$ of both outgoing muons are the same (with opposite direction). Therefore,
we only need the differential
cross section ${\mathrm d}(\delta)\sigma^{\mathrm LO}/{\mathrm d}M{\mathrm d}y{\mathrm d}\phi
{\mathrm d}k_T$ which can be easily obtained \cite{paper}.
Note, that Eq.~(\ref{staterr}) gives the statistical error on the measured asymmetry, 
which is not the same as the theoretical asymmetries shown in Figs.~2 and 4
since usually $\delta\epsilon/\epsilon\neq 1$ (cf. Fig.~3). Fig.~4 shows $A_{TT}$
as a function of the dimuon mass $M$. Just like for $A_{TT}(y)$, the 
expected statistical errors are
sizable for both experiments but {\sc Hera}-$\vec N$ is again in somewhat better shape.
Considering the fact that we assumed a saturation of Eq.~(\ref{soffersinequality}) 
is seems unlikely
that a measurement of $\delta q(x,\mu^2)$ with dimuon production at {\sc Rhic} is possible.
\begin{figure}[htb] 
\vspace{9pt}
\epsfig{file=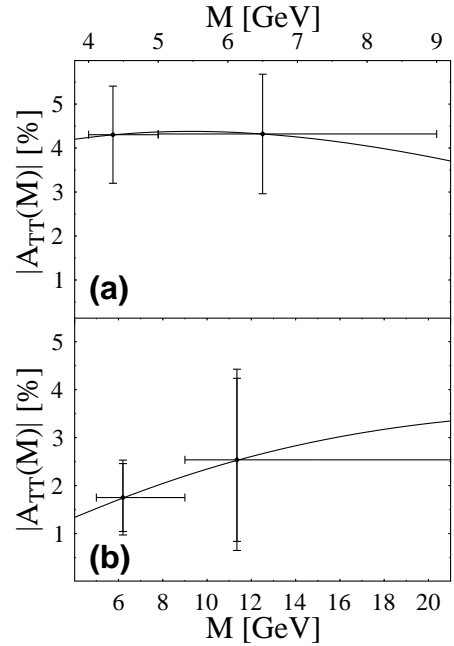,width=6.5cm}
\caption{$A_{TT}(M)$ for (a) {\sc Hera}-$\vec N$ and
(b) {\sc Rhic} (same parameters as in Fig.~2). The two error bars in (b) correspond
to {\sc Phenix} with or without muon detection in the central arms.}
\label{fig4}
\end{figure}

\section*{ACKNOWLEDGEMENTS}
This work was done in collaboration with A.~Sch\"afer, M.~Stratmann and
W.~Vogelsang. The author thanks G.~Bunce, N.~Saito and W.-D.~Nowak
for valuable discussions of experimental issues and E.~Stein for reading
the manuscript.

\end{document}